\documentclass[10pt, a4]{article}
\usepackage[dvipdfmx]{hyperref}
\usepackage[dvipdfmx]{graphicx}
\usepackage{url}
\usepackage{bm}
\usepackage{amsmath}
\usepackage{amsfonts}
\usepackage{multirow}
\usepackage{booktabs}
\usepackage{comment}
\usepackage[top=30truemm,bottom=30truemm,left=25truemm,right=25truemm]{geometry}
\usepackage[ruled]{algorithm2e} 

\SetAlFnt{\small}
\SetAlCapFnt{\small}
\SetAlCapNameFnt{\small}
\SetAlCapHSkip{0pt}
\IncMargin{-\parindent}

\title{F-measure Maximizing Logistic Regression}
\author{
	Masaaki Okabe\thanks{Graduate School of Culture and Information Science, Doshisha University, Tataramiyakodani 1-3, Kyotanabe City, Kyoto, Japan.}
	\and
	Jun Tsuchida\thanks{Faculty of Engineering, Department of Information and Computer Technology, Tokyo University of Science, Niijuku 6-3-1, Katsushika-ku, Tokyo, Japan.}
	\and
	Hiroshi Yadohisa\thanks{Department of Culture and Information Science, Doshisha University, Tataramiyakodani 1-3, Kyotanabe City, Kyoto, Japan} 
	}
\date{}

\begin{document}
\maketitle

\begin{abstract}
	Logistic regression is a widely used method in several fields.
	When applying logistic regression to imbalanced data, for which majority classes dominate over minority classes, all class labels are estimated as ``majority class.''
	In this article, we use an F-measure optimization method to improve the performance of logistic regression applied to imbalanced data.
	While many F-measure optimization methods adopt a ratio of the estimators to approximate the F-measure, the ratio of the estimators tends to have more bias than when the ratio is directly approximated.
	Therefore, we employ an approximate F-measure for estimating the relative density ratio.
	In addition, we define a relative F-measure and approximate the relative F-measure.
	We show an algorithm for a logistic regression weighted approximated relative to the F-measure.
	The experimental results using real world data demonstrated that our proposed method is an efficient algorithm to improve the performance of logistic regression applied to imbalanced data.
\end{abstract}


\section{Introduction}

Class imbalanced data are data for which the ratio of class labels is imbalanced, such as 2:8 or 1:9.
Given the development of the various sophisticated techniques for data acquisition, the amount of imbalanced data has been increasing. 
A classifier trained by imbalanced data tends to predict that all objects belong to the ``majority class.'' This problem occurs because of the assumption of the classifier. 
Many classification methods implicitly assume that the proportion of class labels is 5:5.
To solve this problem, there are three approaches, which can be ultimately considered variations of the weighting method.

The first approach is the sampling method.
To uniformize class balance, we undersample large class objects and oversample small class objects.
The most common method is the synthetic minority over-sampling technique~(SMOTE), proposed by Chawla et al. \cite{chawla2002smote}.
The second approach is the cost-sensitive method.
Under this method, the cost is given for the class misclassifications of each object.
In class imbalanced classification problems, costs are given as reciprocal of the number of classes.
Bahnsen et al. proposed cost-sensitive logistic regression for credit scoring \cite{bahnsen2014example}.

The final approach is changing the misclassification criteria.
Accuracy is the most widely used criteria for classification tasks.
If we use accuracy for a classifier that predicts all subjects as majority class for 1:9 data, it will be 0.9.
In this case, we often use the receiver operating characteristics (ROC) curve and F-measure~\cite{sokolova2006beyond}.
In this method, such as the area under the ROC curve~\cite{bradley1997use} and the F-measure~\cite{Rijsbergen1979}.
The F-measure is commonly used for classification and is defined as the harmonic mean between precision and recall, calculated as the ratio of score and class. However, optimizing the F-measure is often difficult because of its non-convexity. Therefore, many methods of approximating the F-measure have been proposed to solve the optimization problem.

Depending on the approximation method, the F-measure is treated differently. 
One method is using the F-measure as an error function. 
For example, Jansche~\cite{jansche2005maximum} proposed using logistic regression to maximize the approximated F-measure. 
Similarly, Liu et al.~\cite{liu2009regularized} suggested minimizing the approximated log F-measure. 
The model proposed by Liu et al.~\cite{liu2009regularized} is linear, which is similar to a logistic regression model, and uses L1 and L2 regularization. 
Another method is to use an approximated F-measure is as a regularization term. 
For example, Musicant et al.~\cite{musicant2003optimizing} proposed a support vector machine~(SVM) using the F-measure, in which the approximated F-measure is included as a regularization term in the error function of a standard SVM. 
Chinta et al.~\cite{chinta2013optimizing} recommended an SVM for maximizing the F-measure and feature selection simultaneously by using three regularized terms: one based on $L_p$ regularization and the other two based on the F-measure. Ye et al.~\cite{Ye2012optimizing} reported theoretical results for the optimization of the F-measure.

Whether the approximated F-measure is used as an error function or regularization term, its actual approximation is often challenging. Specifically, the F-measure is often approximated by estimating the score and label and, then, taking the ratio of those parameters. However, as pointed out by Huang et al.~\cite{huang2007correcting}, using the ratio of the estimated score and label as an approximation of the F-measure yields more bias than directly approximating the ratio. 

In this article, we thus propose using the relative density ratio~\cite{yamada2013relative} as a direct approximation of the F-measure. 
Additionally, we propose an algorithm for F-measure maximization in the logistic regression model, in which the approximated F-measure we consider as a density ratio is alternatively estimated by using the concept of weighted importance~\cite{shimodaira2000improving}.
Although we focus on logistic regression and binary classification, the concept of our algorithm is useful not only for linear discriminant analyses but also nonlinear discriminant analyses. Moreover, it is easy to expand binary classifications to multi-class and multi-label classifications.

\section{Relevant Background}
\label{sec:rback}
Assume we are given dataset
$\{\bm{x}_i, y_i\}^n_{i=1}$,
$\bm{x}_i \in \mathbb{R}^d$, 
and
$y_i \in \{0, 1\}$.
Let input vector $\bm{x} = (x_1, x_2,..., x_d)^{\top}$ and class label $y$ be binary (i.e., $y$ can be either 1 or 0).

\subsection{Logistic Regression}

Probability $y=1$ can be modeled as follows, given $\bm{x}$:
\begin{align}
	\label{eq:log-p}
	\begin{split}
	P(y=1|\bm{x})
	&= p \\
		&= \frac{1}{1+\exp( - \bm{\beta}^{\top} \bm{x})}, 
	\end{split}
\end{align}
where $\bm{\beta} = (\beta_1, \beta_2,..., \beta_d)^{\top} \in \mathbb{R}^d$.
Assuming $y_i \overset{ind.}{\sim} B(1,p_i)$, the likelihood function can be written as
\begin{align*}
	\ell(\bm{y} | \bm{\beta}) = \prod_{i=1}^{n} p_i^{y_i} (1-p_i)^{1-y_i}.
\end{align*}
We can define the error function by taking the negative logarithm of likelihood, 
which gives a cross-entropy error function $L(\bm{\beta})$ of the following form:
\begin{align*}
	L(\bm{\beta}) = \sum_{i=1}^{n} 
	\left\{
		y_i \log \left( p_i \right)
		+
		(1 - y_i) \log \left( 1- p_i  \right)
	\right\}
.
\end{align*}

Parameter $\bm{\beta}$ in the model is determined so that $L_{\lambda}$, a regularized cross-entropy error function, is minimized.

\begin{align}
	\begin{split}
	L_{\lambda}(\bm{\beta}) = 
	-\sum_{i = 1}^{n}
	\{
		y_i \log(p_i) + (1-y_i) \log(1-p_i)
	\}
	+ \lambda_{\bm{\beta}} ||\bm{\beta}||^2,
	\end{split}
\end{align}
where penalty term $\lambda_{\bm{\beta}} ||\bm{\beta}||^2$ is included for regularization purposes 
and $\lambda_{\bm{\beta}}~(\ge 0)$ denotes the regularization parameter.

\subsection{F-measure}

In a binary classification, an instance is mapped into one of two classes: positive ($y=1$) or negative ($y=0$).
Table~\ref{tab:confusionmatrix} summarizes these outcomes and their
notation.
The number of positive instances is $n_p = \mathrm{TP} + \mathrm{FN}$.
Similarly, the number of negative instances is $n_n = \mathrm{FP} + \mathrm{TN}$.

\begin{table}
	\begin{center}
	\caption{Confusion Matrix}
	\label{tab:confusionmatrix}
	\begin{tabular}{lllll}
		\toprule
		\multicolumn{2}{l}{\multirow{2}{*}{}} & \multicolumn{2}{l}{Predicted Label} & \multirow{2}{*}{Total} \\
			\multicolumn{2}{l}{}                  & 1               & 0               &                        \\
			\midrule
			\multirow{2}{*}{True Label}
			& 1   & TP               & FN               & $n_p$                       \\
			& 0   & FP               & TN               & $n_n$                       \\
			\midrule
			\multicolumn{2}{l}{Total}             & $m_p$                 & $m_n$                 & $n$ \\
    \bottomrule
	\end{tabular}
	\end{center}
\end{table}

We also define precision as follows:
\begin{align}
	\label{eq:precision}
	\mathrm{Precision} = \frac{\mathrm{TP}}{\mathrm{TP}+\mathrm{FP}}
	.
\end{align}
Similarly, recall is defined as
\begin{align}
	\label{eq:recall}
	\mathrm{Recall} = \frac{\mathrm{TP}}{\mathrm{TP}+\mathrm{FN}}
	.
\end{align}
Precision (Equation \ref{eq:precision}) denotes the ratio of the number of true positive (TP) and that of predicted positive instances $m_p$.
Similarly, the recall (Equation \ref{eq:recall}) denotes ratio TP and number of positive instance $n_p$.

The F-measure is defined as
\begin{align}
	F = 2 \left( \frac{1}{\mathrm{Recall}} + \frac{1}{\mathrm{Precision}} \right)^{-1}
	.
\end{align}
This F-measure denotes the harmonic mean of the precision and recall.
Similarly, the $\alpha$-relative F-measure is defined as
\begin{align}
	\label{eq:ReFmeasure}
	\begin{split}
		F_{\alpha}
		&= \left( \alpha \frac{1}{\mathrm{Recall}} + (1 - \alpha) \frac{1}{\mathrm{Precision}} \right)^{-1} \\
		&= \frac{\mathrm{TP}}{\alpha n_p + (1-\alpha) m_p}
	,
	\end{split}
\end{align}
where $0 \le \alpha \le 1$.
The relative F-measure denotes the $\alpha$-weighted harmonic mean of the precision and recall.
When $\alpha = 1/2$, $F_{\alpha}$ is equivalent to $F$.

\section{Proposed Method}

Here, we present the method developed for imbalanced data learning.
We can approximate 
the $\alpha$-relative F-measure (Equation~\ref{eq:ReFmeasure}) with 
the $\alpha$-relative density ratio (Equation~\ref{eq:ardr}).
Weight $w_{\alpha}(\bm{x})$ maps the $\alpha$-relative F-measure.
When $\alpha = 0.5$, the weight $w_{\alpha}(\bm{x})$ is mapped onto the F-measure.


Using our method, 
we can estimate $\bm{\beta}$ 
with a weighted cross-entropy error function to obtain 
the following optimization problem:
\begin{align}
	\begin{split}
\label{eq:pro-beta-obj}
	L_{\bm{\beta}}(\bm{\beta} | \bm{\theta}) = 
	-\sum_{i = 1}^{n}
	w_{\alpha}(p_i)
	\{
		y_i \log(p_i) + (1-y_i) \log(1-p_i)
	\}
	+ \lambda_{\bm{\beta}} ||\bm{\beta}||^2
		,
	\end{split}
\end{align}
where
\begin{align*}
	\begin{split}
		\hat{w}(p_i) = \sum_{\ell=1}^{n}\hat{\bm{\theta}}_i K(p_i,p_{\ell}).
	\end{split}
\end{align*}
The weighted cross-entropy error function (Equation~\ref{eq:pro-beta-obj}) is convex
and its analytical solution
is the global solution.

Assume we are given independent and identically distributed(i.i.d.) samples $\{y_i\}_{i=1}^n$ from a one-dimensional distribution $P_y$ with density $f_{y}(p)$ and i.i.d. samples $\{p_i \}_{i=1}^n$ from the one-dimensional distribution $P_{p}$ with density $f_{p}(p)$:
\begin{align*}
	\{y_i\}_{i=1}^n &\overset{i.i.d.}{\sim} P_y, \\
	\{p_i\}_{i=1}^n &\overset{i.i.d.}{\sim} P_p.
\end{align*}

For $0 \le \alpha \le 1$, let $q_{\alpha}(p)$ be the $\alpha$-mixture density of
$f_{y}(p)$ and $f_p(p)$:
\begin{align*}
	q_{\alpha}(p) &= \alpha f_{y}(p) + (1-\alpha) f_p(p).
\end{align*}
Let weight $w_{\alpha}(p)$ be the $\alpha$-relative weight of $f_{p}(p)$ and $f_{y}(p)$:
\begin{align}
	\label{eq:ardr}
	\begin{split}
	w_{\alpha}(p) 
	&= 
	\frac{f_{y}(p)}{\alpha f_{y}(p) + (1-\alpha) f_p(p)}
	\\
	&=
	\frac{f_{p}(p)}{q_{\alpha}(p)}.
	\end{split}
\end{align}

The relative F-measure is given as
\begin{align*}
	F_{\alpha}
	&= \frac{\mathrm{TP}}{\alpha n_p + (1 - \alpha) m_p}.
\end{align*}
Ideally, because of $\mathrm{FN} = 0$, we can approximate TP by $m_p$, and $F_{\alpha}$ is approximated by:
\begin{align*}
	F_{\alpha}
	\simeq \frac{m_p}{\alpha n_p + (1 - \alpha) m_p}.
\end{align*}
We need to predict the label to obtain $m_p$.
In order to predict the label, we need a prediction model.
We can approximate $m_p$ by $f_p(p)$ and approximate $n_p$ by $f_y(p)$.
Then, $F_{\alpha}$ is approximated by:
\begin{align*}
	\begin{split}
		F_{\alpha}
		&\simeq 
		\frac{f_p(p)}{\alpha f_y(p) + (1-\alpha) f_p(p)} \\
		&= w_{\alpha}(p).
	\end{split}
\end{align*}

Weight $w(p)$ denotes the divergence between the true label and model.
The weight can then be modeled as
\begin{align*}
	w(p; \bm{\theta}) = \sum_{i = 1}^n \theta_i K(p, p_i) ,
\end{align*}
where
$K(p, p_i)$ is a kernel basis function.
In this case, we use the Gaussian kernel:
\begin{align*}
	K(p, p_i) = \exp\bigg( \frac{(p-p_i)^2}{2\sigma} \bigg),
\end{align*}
where $\sigma ( > 0  )$ is the kernel width.
Parameter $\alpha$ adjusts the model as non-parametric or parametric.
When $\alpha = 1$, because of $w(p) = 1$, this method is equivalent to a logistic regression. This is the full parametric model.
On the other hand, when $\alpha = 0$, this is a non-parametric model.
When $\alpha = 0.5$, the F-measure is approximated by $w(p)$.

We can estimate $\bm{\theta}$ via the relative unconstrained least-squares importance fitting(RuLSIF; Yamada et al., 2013\cite{yamada2013relative}), to minimize the following objective function:
\begin{align}
\label{eq:pro-theta-obj}
	L_{\bm{\theta}}(\bm{\theta}|\bm{\beta}) = 
		\frac{1}{2} \bm{\theta}^{\top} \hat{\bm{H}}\bm{\theta} - \hat{\bm{h}}^{\top} \bm{\theta}
		+ \frac{\lambda_{\bm{\theta}}}{2}||\bm{\theta}||^2 ,
\end{align}
where
\begin{align*}
	\hat{H}_{\ell m} 
	&= 
	\frac{\alpha}{n} \sum_{i=1}^n K(p_{i}, p_{\ell}) K(p_{i}, p_{m})+
	\frac{(1-\alpha)}{n} \sum_{i=1}^n K(y_{i}, p_{\ell}) K(y_{i}, p_{m}) ,
	\\
	\hat{h}_{\ell} 
	&= \frac{1}{n} \sum_{i=1}^{n} K(p_{i}, p_{\ell}).
\end{align*}
The tuning parameter $\sigma$ and $\lambda_{\bm{\theta}}$ can be optimized via cross-validation.



\begin{algorithm}[h]
\SetAlgoNoLine
	\KwIn{Imput Data~$\{\bm{x}_i\}^n_{i=1}$ and True Label~$\{y_i\}^n_{i=1}$}
	\KwOut{Parameter Vector~$\bm{\beta}$}

	$\bm{\beta}$ =  minimize CrossEntropy($\{\bm{x}_i\}^n_{i=1}$, $\{y_i\}^n_{i=1}$)\;
	
	\Repeat{$||\bm{\beta} - \tilde{\bm{\beta}}||^2 < \epsilon$}{
		$\tilde{\bm{\beta}} = \bm{\beta}$\;
		$p_i = 1/(1+\exp(-\bm{\beta}^{\top}\bm{x}_i)) \quad (i=1,2,...,n)$\;
		Estimate $\bm{\theta}$ via RuLSIF\;
		$\bm{\beta}$ = minimize WeightedCrossEntropy($\bm{\theta}$, $\{\bm{x}_i\}^n_{i=1}$, $\{y_i\}^n_{i=1}$)\;
		}
\caption{Estimation Algorithm}
\label{alg:prop}
\end{algorithm}

\section{Experiment}
We compare the proposed method, whose $\alpha$ was set to 0.5, 
with 
L2 regularized logistic regression, 
L2 regularized logistic regression via SMOTE\cite{chawla2002smote},
L2 regularized logistic regression via cost-sensitive method \cite{bahnsen2014example}
and Janche's model 
by the area under the curve~(AUC) of the testing data.
AUC is defined as follows:
\begin{align}
\text{AUC} = \frac{\sum_{i=1}^{n_p}\sum_{j=1}^{n_n}I(p_i >p_j) }{n_p n_n},
\end{align}
where $I(\cdot)$ is an indicator function. 

To tune the hyper-parameters of the proposed method and L2 regularized logistic regression, we used grid search. 
Specifically, we set the range of penalty terms $\lambda_{\bm{\theta}}$ and $\lambda_{\bm{\beta}}$ to $\{0.01, 0.1, 1, 10\}$, the range of $\sigma$ for the Gaussian kernel to $\{0.01, 0.1, 1, 10\}$, and the range of the number of variables to $\{1, 2, \cdots, d\}$. 
We selected the best AUC performance hyper-parameters by five-fold cross-validation.

\subsection{Real Data Example}

Table~\ref{tab:data} gives an overview of the datasets. 
We used two types of datasets to evaluate performance: one with a data ratio of 2:1 between majority and minority class labels and one with a ratio of 9:1.
The Yeast\footnote{Because the Yeast dataset is not binary, we replaced the ME2 class to a positive class and other classes to negative ones.} dataset was taken from the UCI Repository of Machine Learning Databases at
\url{http://www.ics.uci.edu/~mlearn/MLRepository.html} and the Oil dataset from Open ML (\url{https://openml.org/}).

\begin{table}
	\begin{center}
	\caption{Overview of datasets. The minority class is a positive class. The ratio denotes the minority class ratio.}
  \label{tab:data}
  \begin{tabular}{ccccc|c}
    \toprule
	  Dataset & Size & Features & Positive & Negative & Positive Case Ratio\\
    \midrule
	  Oil & 937 & 48 & 41 & 896 & 0.0438 \\
	  Yeast & 1484 & 9 & 51 & 1433 & 0.0344 \\
	  \bottomrule
  \end{tabular}
\end{center}
\end{table}

We first eliminated objects with NA observations.
Then, the datasets were randomly divided into training and testing data. 
The ratio of the number of objects in the training and testing datasets was 7:3. The data were then normalized with a zero mean and standard deviation. 

Tables~\ref{tab:resreal-auc}, \ref{tab:resreal-f1}, and \ref{tab:resreal-acc} respectively show the mean and standard deviation of AUC score, F-measures, and accuracy.
From table~\ref{tab:resreal-auc}, SMOTE has good performance in terms of AUC.
From table~\ref{tab:resreal-f1}, the proposed method shows a good performance by the F-measure.
Finally, from table~\ref{tab:resreal-acc}, logistic regression has the highest accuracy score for the Oil dataset.
This proves the proposed method has the highest F-measure, meaning we could obtain better discriminant result by using the F-measure.


\begin{table}[!tbp]
\begin{center}
\caption{AUC of the real data example \label{tab:resreal-auc}}
\begin{tabular}{lrrrrr}
\hline\hline
\multicolumn{1}{l}{}&\multicolumn{1}{c}{Proposed}&\multicolumn{1}{c}{Logistic}&\multicolumn{1}{c}{Cost-sensitive}&\multicolumn{1}{c}{Janche}&\multicolumn{1}{c}{SMOTE}\tabularnewline
\hline
{\bfseries Oil}&&&&&\tabularnewline
	~~mean&$0.774$&$0.757$&$0.625$&$0.568$&${\bf 0.831}$\tabularnewline
	~~s.d.&$0.144$&$0.123$&$0.104$&${\bf 0.036}$&$0.116$\tabularnewline
\hline
{\bfseries Yeast}&&&&&\tabularnewline
	~~mean&$0.610$&$0.568$&$0.555$&$0.572$&${\bf 0.800}$\tabularnewline
	~~s.d.&$0.118$&$0.107$&$0.092$&${\bf 0.064}$&$0.078$\tabularnewline
\hline
\end{tabular}\end{center}
\end{table}

\begin{table}[!tbp]
\begin{center}
\caption{F-measure of the real data example \label{tab:resreal-f1}}
\begin{tabular}{lrrrrr}
\hline\hline
\multicolumn{1}{l}{}&\multicolumn{1}{c}{Proposed}&\multicolumn{1}{c}{Logistic}&\multicolumn{1}{c}{Cost-sensitive}&\multicolumn{1}{c}{Janche}&\multicolumn{1}{c}{SMOTE}\tabularnewline
\hline
{\bfseries Oil}&&&&&\tabularnewline
	~~mean&${\bf 0.596}$&$0.582$&$0.121$&$0.237$&$0.489$\tabularnewline
	~~s.d.&$0.266$&$0.242$&${\bf 0.056}$&$0.107$&$0.222$\tabularnewline
\hline
{\bfseries Yeast}&&&&&\tabularnewline
	~~mean&${\bf 0.282}$&$0.187$&$0.103$&$0.228$&$0.258$\tabularnewline
	~~s.d.&$0.266$&$0.239$&${\bf 0.071}$&$0.149$&$0.106$\tabularnewline
\hline
\end{tabular}\end{center}
\end{table}

\begin{table}[!tbp]
\begin{center}
\caption{Accuracy of the real data example \label{tab:resreal-acc}}
\begin{tabular}{lrrrrr}
\hline\hline
\multicolumn{1}{l}{}&\multicolumn{1}{c}{Proposed}&\multicolumn{1}{c}{Logistic}&\multicolumn{1}{c}{Cost-sensitive}&\multicolumn{1}{c}{Janche}&\multicolumn{1}{c}{SMOTE}\tabularnewline
\hline
{\bfseries Oil}&&&&&\tabularnewline
	~~mean&$0.966$&${\bf 0.970}$&$0.551$&$0.728$&$0.935$\tabularnewline
	~~s.d.&$0.023$&${\bf 0.017}$&$0.200$&$0.101$&$0.028$\tabularnewline
\hline
{\bfseries Yeast}&&&&&\tabularnewline
	~~mean&${\bf 0.968}$&$0.967$&$0.253$&$0.653$&$0.854$\tabularnewline
	~~s.d.&${\bf 0.012}$&$0.014$&$0.364$&$0.119$&$0.036$\tabularnewline
\hline
\end{tabular}\end{center}
\end{table}

\subsection{Numerical Example}
We created the synthetic data as follows.

\begin{itemize}
	\item $\bm{X}_i \overset{i.i.d.}{\sim} N_d(\bm{0}, \bm{I}_d)$
	\item $p_{i} = \frac{1}{1+\exp(-\bm{\beta}^{\top}\bm{X}_i)}$
	\item $Y_{i} \overset{ind.}{\sim} B(1, p_{i})$
\end{itemize}

Table~\ref{tab:tbeta} shows the setting of the true $\bm{\beta}$ and positive case ratio for each dataset.
Test data were also created under these settings.

\begin{table}
	\begin{center}
		\caption{True $\bm{\beta}$ and positive case ratio}
  \label{tab:tbeta}
  \begin{tabular}{llc}
    \toprule
	   & True $\bm{\beta}$ & Positive case ratio\\
    \midrule
	  Setting 1 & $(-1,1,0,0,0,0,0,0,0,0)   $ & 0.29\\
	  Setting 2 & $(-1,0,-1,-1,1,-2,0,0,0,0)$ & 0.37\\
	  Setting 3 & $(1,0,0,0,0,0,0,-1,2,0)   $ & 0.64\\
	  Setting 4 & $(0,0,0,-1,2,0,0,0,0,0)   $ & 0.50\\
	  Setting 5 & $(-4,0,0,0,2,0,0,0,0,0)   $ & 0.06\\
	  Setting 6 & $(4,0,0,3,0,0,0,0,0,0)    $ & 0.88\\
	  \bottomrule
  \end{tabular}
\end{center}
\end{table}

Table~\ref{tab:resnume-auc} shows the AUCs for the numerical example.
SMOTE has the best performance under settings 3, 5, and 6. On the other hand, the proposed method's best performance is under settings 1, 2, and 4. 
For class imbalanced data, SMOTE shows good performance from the perspective of AUC.
Table~\ref{tab:resnume-f1} presents the F-measures for the numerical example.
The proposed method has the best performance of all settings.
As such, for class imbalanced data, the proposed method shows good performance from the perspective of using the F-measure.
Table~\ref{tab:resnume-acc} gives the accuracy for the numerical example, with the proposed method and logistic regression having the best performances.

However, from Tables~\ref{tab:resnume-auc} and ~\ref{tab:resnume-f1}, the proposed method is not better at estimating the probability of $\hat{p}$ as it has better performance for discrimination, but not AUC scores.

\begin{table}[!tbp]
\begin{center}
\caption{AUC of numerical example \label{tab:resnume-auc}}
\begin{tabular}{lrrrrr}
\hline\hline
\multicolumn{1}{l}{}&\multicolumn{1}{c}{Proposed}&\multicolumn{1}{c}{Logistic}&\multicolumn{1}{c}{Cost-sensitive}&\multicolumn{1}{c}{Janche}&\multicolumn{1}{c}{SMOTE}\tabularnewline
\hline
{\bfseries Setting 1}&&&&&\tabularnewline
	~~mean&${\bf 0.667}$&$0.640$&$0.518$&$0.614$&$0.654$\tabularnewline
	~~s.d.&$0.058$&$0.058$&${\bf 0.031}$&$0.052$&$0.061$\tabularnewline
\hline
{\bfseries Setting 2}&&&&&\tabularnewline
	~~mean&${\bf 0.818}$&$0.804$&$0.526$&$0.745$&$0.811$\tabularnewline
	~~s.d.&${\bf 0.043}$&$0.044$&$0.047$&$0.052$&$0.046$\tabularnewline
\hline
{\bfseries Setting 3}&&&&&\tabularnewline
	~~mean&$0.803$&$0.793$&$0.532$&$0.787$&${\bf 0.805}$\tabularnewline
	~~s.d.&$0.044$&$0.044$&$0.055$&$0.052$&${\bf 0.041}$\tabularnewline
\hline
{\bfseries Setting 4}&&&&&\tabularnewline
	~~mean&${\bf 0.805}$&$0.797$&$0.510$&$0.758$&$0.795$\tabularnewline
	~~s.d.&$0.036$&$0.037$&${\bf 0.019}$&$0.050$&$0.036$\tabularnewline
\hline
{\bfseries Setting 5}&&&&&\tabularnewline
	~~mean&$0.696$&$0.625$&$0.532$&$0.562$&${\bf 0.714}$\tabularnewline
	~~s.d.&$0.104$&$0.094$&${\bf 0.063}$&${\bf 0.063}$&$0.108$\tabularnewline
\hline
{\bfseries Setting 6}&&&&&\tabularnewline
	~~mean&$0.805$&$0.772$&$0.527$&$0.682$&${\bf 0.831}$\tabularnewline
	~~s.d.&$0.085$&$0.082$&${\bf 0.062}$&$0.162$&$0.071$\tabularnewline
\hline
\end{tabular}\end{center}
\end{table}

\begin{table}[!tbp]
\begin{center}
\caption{F-measure of numerical example \label{tab:resnume-f1}}
\begin{tabular}{lrrrrr}
\hline\hline
\multicolumn{1}{l}{}&\multicolumn{1}{c}{Proposed}&\multicolumn{1}{c}{Logistic}&\multicolumn{1}{c}{Cost-sensitive}&\multicolumn{1}{c}{Janche}&\multicolumn{1}{c}{SMOTE}\tabularnewline
\hline
{\bfseries Setting 1}&&&&&\tabularnewline
	~~mean&${\bf 0.515}$&$0.463$&$0.390$&$0.503$&$0.511$\tabularnewline
	~~s.d.&$0.099$&$0.109$&$0.143$&${\bf 0.076}$&$0.088$\tabularnewline
\hline
{\bfseries Setting 2}&&&&&\tabularnewline
	~~mean&${\bf 0.767}$&$0.749$&$0.518$&$0.695$&$0.757$\tabularnewline
	~~s.d.&${\bf 0.059}$&$0.062$&$0.109$&$0.071$&$0.065$\tabularnewline
\hline
{\bfseries Setting 3}&&&&&\tabularnewline
	~~mean&${\bf 0.858}$&$0.854$&$0.665$&$0.839$&$0.839$\tabularnewline
	~~s.d.&$0.032$&${\bf 0.031}$&$0.244$&$0.040$&$0.035$\tabularnewline
\hline
{\bfseries Setting 4}&&&&&\tabularnewline
	~~mean&${\bf 0.804}$&$0.796$&$0.605$&$0.767$&$0.794$\tabularnewline
	~~s.d.&${\bf 0.039}$&$0.042$&$0.171$&$0.047$&$0.041$\tabularnewline
\hline
{\bfseries Setting 5}&&&&&\tabularnewline
	~~mean&${\bf 0.376}$&$0.295$&$0.117$&$0.218$&$0.309$\tabularnewline
	~~s.d.&$0.181$&$0.192$&${\bf 0.085}$&$0.130$&$0.145$\tabularnewline
\hline
{\bfseries Setting 6}&&&&&\tabularnewline
	~~mean&${\bf 0.956}$&$0.953$&$0.782$&$0.720$&$0.933$\tabularnewline
	~~s.d.&${\bf 0.017}$&${\bf 0.017}$&$0.315$&$0.343$&$0.023$\tabularnewline
\hline
\end{tabular}\end{center}
\end{table}

\begin{table}[!tbp]
\begin{center}
\caption{Accuracy of numerical example \label{tab:resnume-acc}}
\begin{tabular}{lrrrrr}
\hline\hline
\multicolumn{1}{l}{}&\multicolumn{1}{c}{Proposed}&\multicolumn{1}{c}{Logistic}&\multicolumn{1}{c}{Cost-sensitive}&\multicolumn{1}{c}{Janche}&\multicolumn{1}{c}{SMOTE}\tabularnewline
\hline
{\bfseries Setting 1}&&&&&\tabularnewline
	~~mean&${\bf 0.756}$&$0.747$&$0.639$&$0.559$&$0.681$\tabularnewline
	~~s.d.&${\bf 0.049}$&$0.050$&$0.151$&$0.087$&$0.057$\tabularnewline
\hline
{\bfseries Setting 2}&&&&&\tabularnewline
	~~mean&${\bf 0.835}$&$0.824$&$0.616$&$0.706$&$0.816$\tabularnewline
	~~s.d.&${\bf 0.037}$&${\bf 0.037}$&$0.078$&$0.070$&$0.043$\tabularnewline
\hline
{\bfseries Setting 3}&&&&&\tabularnewline
	~~mean&${\bf 0.819}$&$0.813$&$0.601$&$0.773$&$0.804$\tabularnewline
	~~s.d.&${\bf 0.034}$&$0.035$&$0.102$&$0.056$&$0.039$\tabularnewline
\hline
{\bfseries Setting 4}&&&&&\tabularnewline
	~~mean&${\bf 0.805}$&$0.798$&$0.526$&$0.725$&$0.795$\tabularnewline
	~~s.d.&$0.037$&$0.038$&${\bf 0.033}$&$0.055$&$0.037$\tabularnewline
\hline
{\bfseries Setting 5}&&&&&\tabularnewline
	~~mean&$0.937$&${\bf 0.938}$&$0.735$&$0.630$&$0.870$\tabularnewline
	~~s.d.&$0.024$&${\bf 0.023}$&$0.315$&$0.156$&$0.041$\tabularnewline
\hline
{\bfseries Setting 6}&&&&&\tabularnewline
	~~mean&${\bf 0.922}$&$0.917$&$0.741$&$0.884$&$0.886$\tabularnewline
	~~s.d.&${\bf 0.028}$&$0.029$&$0.268$&$0.036$&$0.037$\tabularnewline
\hline
\end{tabular}\end{center}
\end{table}


\section{Conclusions}

In this article, we proposed an F-measure maximizing logistic regression using the relative density ratio.
We show that the proposed method has better performance than the others from the perspective of the F-measure in the class imbalanced real data example.
In this result, the proposed method is useful in the case of a discriminant.
On the other hand, 
the AUC of the proposed method is not higher than that of SMOTE.
The reason is that the weights of object near the decision boundary are higher than in the other methods.
That is, the order of estimated probability is unstable near the decision boundary.

Our future work will involve developing this theoretical approach. There are three problems to be tackled.
First, the theoretical predictive performance of our model is unclear. In the experiment, our method is used on imbalanced data. However, it is unclear asymptotic properties.
Second, the performance of the approximation to the relative F-measure should be investigated. In the experiment, the prediction performance of our model is effective from the view of AUC. However, we should discuss the performance of the approximation as well.
Finally, we should show the appropriateness of the relationship of the density ratio of $p$ and $\hat{p}$, and of the F-measure.
Overall, we obtained better discriminant performance from using the density ratio experimentally, but did not obtain theoretically relevant results.

\bibliography{bib_list.bib}

\end{document}